\documentclass[12pt,aps,prd,preprint,showkeys]{revtex4}
\usepackage[utf8]{inputenc}
\usepackage{amsmath}
\usepackage{amsfonts}
\usepackage{amssymb}
\usepackage{graphicx}
\usepackage{csquotes}

\begin{document}

\title{Synchrotron radiation from cosmic string wakes. }
\author{Dilip Kumar, Soumen Nayak and Soma Sanyal}
\affiliation{School of Physics, University of Hyderabad, Gachibowli, Hyderabad, India 500046}


\begin{abstract}

{Magnetic fields can be generated in cosmic string wakes due to the Biermann mechanism in the presence of neutrino inhomogeneities. As the cosmic string moves through the plasma the small magnetic field is amplified by the turbulence in the plasma. Relativistic charged particles which cross the magnetized wake of a cosmic string will therefore emit synchrotron radiation. The opening 
angle of the cosmic string is very small and so the wake appears like a thin sheet. Assuming a homogeneous magnetic field in the wake of the string, we obtain the synchrotron emission from non thermal relativistic electrons in the wake of the string. The emitted radiation has a broad peak and is over a wide range of frequency. We show that the spectrum can be mapped to some of the unknown sources in different ranges of the current available catalogues. } 

\end{abstract}


\maketitle

\section{Introduction}

Synchrotron radiation has been studied for a very long time to identify and locate various non-luminous astrophysical objects \cite{blackhole2021} in our universe. It is the radiation emitted from relativistic electrons moving in the magnetic field close to an astrophysical object. One of the most elusive objects which are currently being 
sought using various methods are the cosmic strings. These strings are 
produced during symmetry breaking phase transitions in the early universe \cite{csreview}. They are essentially defects in one dimension with a conical spacetime. These strings can give rise to interesting phenomenon in the early universe. The conical space around the string leads to the formation of wakes behind them \cite{deruelle}. In relativistic fluid flows, both strong shocks as well as weak shocks are generated \cite{hiscock,vollick,beresnyak} in these wakes. There are many kinds of cosmic strings depending on the nature of the phase transition. Out of all these, the superconducting cosmic string carry electric current and can generate magnetic fields in their wake \cite{dimopoulos}. So, the spectrum of synchrotron radiation from electrons moving close to a superconducting string has been studied in detail \cite{rogozin}. Recently, it has been shown that magnetic fields can be formed in shocks of wakes of non-superconducting strings by the Biermann battery mechanism \cite{sovan}. The Biermann battery mechanism generates a magnetic field whenever the temperature gradient and the density gradient in an inhomogeneous plasma are not aligned to each other \cite{schoeffler}. Once, the magnetic field is generated, it gets enhanced by the turbulence in the wake region. As the cosmic string moves in the plasma, electrons crossing these magnetized wakes will emit synchrotron radiation. The energy spectrum of these electrons along with other signatures may help in identifying these cosmic strings in the current universe. Currently, there are various attempts to predict the signature of these cosmic string from the 21 cm observational data \cite{maibach}. Other methods of identifying cosmic strings are also being pursued. In this article, we are proposing yet another method of identifying cosmic string wakes by using electromagnetic radiation. 

The study of diffusive synchrotron radiation has already yielded several interesting results. The spectrum of the radiation depends on the correlation length of the magnetic field through which the electrons travel. For a uniform large scale magnetic field the spectrum is referred to as the synchrotron spectrum while for small scale random magnetic field the radiation is often referred to as the \enquote{jitter} radiation \cite{medvedev,fleishman}. A theory has also been developed for the small scale jitter radiation. The theory gives rise to several power law asymptotes which smoothly change as the frequency changes. This leads to the fact that a wide range of frequencies are now covered by the synchrotron radiation.    

 In recent times, numerical simulations have been developed to study the microphysics of non-relativistic as well as relativistic shocks.  For cosmic strings too, there have been several studies of unmagnetized shocks \cite{stebbins,sornborger,ponce}. The wakes of cosmic strings are planar structures and the shocks are generated in these planar wakes. In a previous work, we have studied the evolution of a magnetic field in the planer wake of a cosmic string \cite{nayak}. Cosmic strings can have magnetic fields in their wakes due to various mechanisms. One mechanism that has been mentioned before is the Biermann battery mechanism. There are alternative mechanisms which involve the generation of vorticity in the plasma. Two cosmic strings moving past one another can generate vortices between them \cite{vachaspati}. Since ions and electrons do not have the same mass, a magnetic field can be generated by the differential angular momentum between the ions and electrons  via the Harrison mechanism \cite{harrison}. Other ways of generating vorticity have also been explored. Once the field is generated in the wake, it may decay but it will not dissipate completely, so the electrons moving across these wakes will be moving through a magnetic field. This will  lead to the emission of synchrotron radiation. Previously, only the synchrotron radiation coming from wakes of superconducting strings have been studied in the literature. In this article we will focus on Abelian Higgs strings.

 Unlike superconducting strings these do not have a current flowing through the length of the string. The magnetic field in these wakes are generated by the Biermann mechanism \cite{sovan}. The Biermann mechanism generates oppositely directed magnetic fields on the different sides of the planar wake. The wake has a triangular structure,  being narrow close to the string and becoming wider as the distance from the string increases. Multiple shocks are generated in the magnetized wakes \cite{nayak} of cosmic strings. Electrons will move across the wake as the string moves through the plasma and will be accelerated by the shocks in the magnetized plasma. We obtain the spectrum of synchrotron radiation emitted by these electrons in the cosmic string wake.  We find that, the overall spectrum  has a broad peak. Due to the large range of frequencies that it spans, the radiation should be detected by the current all sky surveys. We discuss how this can be another signature for cosmic string wakes in addition to the ones that are already being pursued.

In section 2, we distinguish between superconducting cosmic strings and the Abelian Higgs cosmic strings and their wakes in the intergalactic medium.  In section 3, we discuss how electrons are accelerated in the shocks of cosmic strings. In section 4, we find the synchrotron radiation emitted by the electrons moving in the magnetized wake of the cosmic strings. In section 5, we present our results and obtain the spectrum of synchrotron radiation. In section 6, we present a comparison between the synchrotron radiation from superconducting strings and the spectrum that we have obtained. Finally in Section 7, we present the summary and conclusions of the paper.

\section{Cosmic string wake in the intergalactic medium} 

Cosmic string wake forms as a long cosmic string moves through the intergalactic medium. The Abelian Higgs string which is formed due to the breaking of the Abelian Higgs symmetry is one
of the simplest cosmic string to be studied, having a minimal number of degrees of freedom. The major parameter defining these strings is the deficit angle of the conical space time around them. This is related to the symmetry breaking scale and the ratio of the scalar and the vector couplings of the Lagrangian. The ratio of the couplings is denoted by the quantity $\xi = \lambda/e^2$, where $\lambda$ is the scalar self coupling and $e$ is the gauge coupling present in the Lagrangian for the Abelian Higgs theory.

As mentioned before, specific Lagrangians can also give rise to a superconducting string. The major distinguishing factors 
of these superconducting strings is that they are current carrying strings. These currents can
be due to localized charged condensates or fermionic zero modes which carry electric charge.  Lagrangian models, including the electroweak model, may give rise to such current 
carrying cosmic strings \cite{hill,ppeter}. This current ($i$) is the primary reason for the generation of magnetic fields in the wake of a superconducting string and it has been shown that the generation of synchrotron radiation in the wake of these strings will also depend on the current carried by the superconducting cosmic string \cite{rogozin}. The synchrotron radiation in the planar wakes of these superconducting strings do not depend on the length of the strings. In our case however, the Abelian Higgs string is not a current carrying cosmic string, hence the generated magnetic field and the subsequent calculation of synchrotron radiation does not depend on any string parameter except the deficit angle which is related to the symmetry breaking scale of the cosmic string. We will briefly discuss the wakes generated by Abelian Higgs strings due to the presence of the deficit angle.

The wakes due to long strings occur due to planar accretion of matter \cite{sornborger}. They are therefore often treated as two dimensional structures instead of three.  Detailed numerical studies have been carried out for wakes generated by long strings which do not have any current flowing through them. A string moving at a time $t_{i}$ will generate a wake whose dimensions are given by $c_1 t_i \times t_i v_s \gamma_s \times \delta \theta  t_i v_s \gamma_s $, where $v_s$ is the velocity of the string and $\gamma_s$ is the relativistic Lorentz factor and $c_1$ is a constant of order one \cite{branden21}. Out of these three dimensions, one dimension is much much smaller than the other two due to the fact that it is dependent on the symmetry breaking scale ($G \mu \approx 10^{-7}$). Thus the wake is usually treated as a planar structure where only two dimensions are important.  It has been shown that the wake is  wide for a hot gas tightly coupled to radiation but becomes quite narrow for the case of a hydrogen gas \cite{ponce}. Since the intergalactic medium consists of mostly baryonic and leptonic matter, hence the wake caused by a moving cosmic string would be closer to the case of the hydrogen gas. Thus, the wake would be a narrow stream of relativistic particles moving with relativistic velocities. As the velocity of the string reaches supersonic velocities, shock waves are generated behind the cosmic string. The shocks are generated in the planar structure of the wake and the different particles present in the plasma move across these shocks. We will give some details of the cosmic string shocks previously studied by various other 
authors.    

Cosmic string shock waves have been discussed for the high temperature plasma of the very early universe. The relativistic motion of the string as well as the finite temperature of the medium has been taken into account in these studies \cite{stebbins,rees,deruelle}. These shocks occur before the start of recombination. Realistic studies of shocks also include the the
possibility of ionization and interaction with the background
radiation \cite{ponce}. The velocity of the shock would depend on the nature 
of the plasma and it's Equation of State (EoS). In the post recombination era, the universe is matter dominated. Assuming that the interaction between the matter and radiation is minimal, the sound speed at those temperatures is given by, 
\begin{equation}
c_s^2 = \frac{\Gamma_m p_1}{\rho_1 + p_1} \label{c_s}
\end{equation} 
Here, $\Gamma_m $ is the ratio of the specific heats of a radiation gas and can be considered to be equal to $5/3$, while $p_1$ and $\rho_1 $ are the pressure and density in the pre-shock region respectively. Detailed studies of shocks in these plasmas have been carried out \cite{ponce}. The wakes usually have different values of density, pressure and temperature in the pre-shock and the post-shock regions. The shock which is similar to a front moves with a velocity known as the shock velocity. Generally, this shock velocity depends both on the pre-shock and post-shock density and pressure. It has been shown that, in the case, where the specific heat ratio remains approximately the same in the pre-shock and the post-shock regime, one can obtain the velocity of the shock wave as 
\begin{equation}
v_{sh} = \frac{1}{4} [(\Gamma_m+ 1)^2 u^2 + 16 c_s^2]^{1/2} - \frac{(3-\Gamma_m)}{4} u  \label{v_sh}
\end{equation}  
Here $u = v_s \delta \theta (1-v_s^2)^{-1/2}$ , $\delta \theta$ is the angle of deflection of the particles due to the cosmic string metric and $v_s$ is the velocity of the string. Therefore, the shock velocities can go as high as $0.9 c$ \cite{delaney}. The matter inside the wake region will be bounded by the shock fronts. 
Though the gas as a whole can be considered to be static, the internal energy of the particles would be a sum of the thermal and kinetic energies of the particles which are moving with velocities depending upon the density and temperature of the gas \cite{rees}. 
These wakes which have been studied previously in the case of the cosmic strings are hydrodynamic shock waves. The presence of the magnetic field will lead to magnetohydrodynamic (MHD) shock waves.     
Recently MHD shock waves have been studied numerically for cosmic string wakes \cite{nayak}.

Though similar to hydrodynamic shocks, MHD shocks have the added property of accelerating electrons in the region bounded by the shock waves. Electron acceleration from various shock waves have been studied in detail for other situations\cite{galeev}. Both sub-relativistic shock waves as well as ultra relativistic shock waves are known to generate accelerated electrons which subsequently emit synchrotron radiation. The plasma we are considering will have both leptons and baryons. In this work, we show how the electrons moving across the wake are accelerated due to the presence of the shock waves at the boundary of the wake.  
The accelerated electrons will in turn emit synchrotron radiation.
This is discussed in Section III, before that we will discuss some basic properties of hydrodynamic shocks that have previously been studied for cosmic string wakes.

One of the important reasons for electron acceleration in the wake region, bounded by the shocks is the inhomogeneity in the number density of the particles in the wake region. Due to the
conical metric, all the particles will have a net angular momentum close to the cosmic string 
wake. This arises due to the velocity perturbation felt by a particle as it moves past the 
cosmic string. The velocity perturbation is given by $\delta v \sim \delta \theta v_s \gamma_s $.  The Fermi distribution of electrons (or neutrinos) about the cosmic string
is given by, \cite{sovan} 
\begin{equation}
f(E, l_z, \chi) = \left [exp\left(\frac{E - l_z \Omega - \mu \chi}{T} \right) + 1 \right]^{-1} \label{fermi}
\end{equation}
where $\Omega$ is the angular velocity, $\mu$ is the chemical
potential of the particles being considered, and $l_z$ is the projection of the particle's total angular momentum  on the direction of $\Omega$ , and $E$ is the energy of the particles. The factor $\chi$ takes on the values $1$ or $-1$ depending on whether the particles are more than the antiparticles in the plasma.

In the very early universe, depending on their energy and angular momentum, particles appear to cluster closer to the string due to the presence of bounded orbits \cite{saha}. Due to the asymmetry amongst the right and left handed neutrinos, a neutrino current is generated. This neutrino current is oscillating in nature and the number density of the neutrinos depends on their distance from the core of the cosmic string \cite{sovan}. The detailed calculations have previously been presented in ref.\cite{sovan}. The neutrinos and electrons interact due to the ponderomotive force. The
number density of neutrinos being oscillatory in nature, the electrons on interacting with
them will cease to have a uniform distribution. They will be pushed into the regions where 
there is an under density of neutrinos. So, the wake structure will not have a uniform density 
of electrons and neutrinos. Their number densities will be inhomogeneous in nature. 

These are however small scale inhomogeneities. The wake itself has a large scale overdensity as we have mentioned in the initial part. This is due to the velocity perturbation that is felt by all the matter particles irrespective of whether they are neutrinos or electrons. The overall temperature  in the pre-shock and post-shock region is determined by this larger overdensity.  
The temperature in the post-shock regime is given by, 
\begin{equation}
T_2 = \frac{p_2 n_1}{p_1 n_2} T_1 \label{T_2}
\end{equation}   
Here the suffix $1$ is used for the pre-shock quantities, while the suffix $2$ is used for the post-shock quantities. The overdensity in the cosmic string wake is given by \cite{layek}
\begin{equation}
\frac{\delta \rho}{\rho } = \frac{16 \pi G \mu v_s^2}{3 \sqrt{v_s^2 - v_c^2}} \label{overdensity}
\end{equation} 
where $\rho$ is the average plasma density and $\delta \rho$ is the amount of the overdensity. $v_c$ is the sound speed in the plasma. The opening angle of the wedge is related to the velocities as 
\begin{equation}
sin~ \theta \approx \frac{v_c}{v_s} \label{wedge angle}
\end{equation}
The angle $\theta \approx 8 \pi G \mu$ is very small, so the velocity of the cosmic string is close to that of light. The string may also move at ultra relativistic velocities, however most simulations consider only relativistic speeds for a cosmic string network at lower values of redshift \cite{sornborger2}. 

In the next section, we describe how the electrons are accelerated in shocks of cosmic string wakes. Later, we will use the values of these shock accelerated electrons to determine the spectrum of the synchrotron radiation.

\section{Electron acceleration in shocks of cosmic string wakes}

There are several studies of cosmic string wakes in the literature. Numerical studies have shown multiple shock waves in the magnetized wake of a cosmic string \cite{nayak}. Due to the presence of the deficit angle, the shocks formed would be quasi perpendicular to the direction of motion 
of the string. It has also been shown that in cosmic string wakes, the baryons tend to cluster 
towards the center of the wake \cite{sornborger}. This makes the overall baryon density quite inhomogeneous in the wake.  
The presence of a magnetic field in the wake
would mean that both ions and electrons are accelerated by the magnetic field. In such cases, it 
has been seen that the accelerated ions are reflected by the strong shock front generated in 
the plasma. A detailed study in ref. \cite{galeev} shows that the electrons get scattered by the 
reflected ions repeatedly. The repeated reflections accelerate the electrons and the energy spectrum of the 
subsequent electron distribution can be represented by a power law with an exponent of $2$ (approximately). While the reflected ions tend to accelerate the electrons, the Landau damping
of the waves would lead to a decrease in their energy. As has been shown in ref \cite{galeev}, 
a balance has to be generated between the excitation and the loss. Since at greater wavelengths, it is difficult to reach a balance, the energy spectrum appears to have a jet like structure. The efficiency of the acceleration of the electrons is characterized by 
the pressure difference across the shock as well as the velocity 
of the reflected ions. The pressure difference across the shock wave can be quite significant in the case of a cosmic string. 
Strong shocks are known to be formed in the wakes of cosmic strings after recombination \cite{sornborger, vachaspati}. 

In this case, as the ions move away from the center of the wake they are typically reflected back due to the collision with the strong shock front. The electrons are scattered of these reflected ions and a significant fraction of the energy is transferred to them. The electrons are thus accelerated to very high energies. 
Following the analysis in ref \cite{galeev}, we therefore consider 
a power law distribution of the electrons in the post-shock region with electron Lorentz factors $\gamma_{min} < \gamma < \gamma_{max}$. The range of the Lorentz factor comes due to the fact that the energy is continuously injected into the  electron distribution in the comoving reference frame.  The minimum Lorentz factor would be given by the minimum velocity with which the electrons are swept inside the wake of the cosmic string. If the electrons are not reflected or accelerated by the ions reflected from shocks, they would have the same $\gamma$ as the Lorentz factor of the cosmic string. Hence the minimum possible Lorentz factor of the electrons in the distribution would be one (since the 
velocity of the string is approximately $0.5 c$ \cite{sornborger2}). 

For the maximum Lorentz factor $\gamma_{max}$, we need to take care of not only the energy being injected in the electron distribution 
but also the energy being taken away by various processes. As has
been already mentioned, there is a fine balance between the energy
injection and the energy decay which gives us the power law spectrum. Hence one should look at the dominant mechanism at which these electrons lose their energy. If the synchrotron mechanism is 
the dominant mechanism for energy loss then the maximum electron Lorentz factor is set primarily by the rate at which the electron
accelerates. To find out if the synchrotron emission is the dominant method for 
energy loss, we need to calculate the cooling electron Lorentz factor \enquote{$\gamma_c$}. This factor comes from the adiabatic losses in the electron distribution in the shock as it spreads out in the post-shock region. In the post- shock region, the string has already moved away and there is a decelerating effect on the electrons. The energy loss can be studied taking into consideration the $\gamma$ factor of the decelerating electrons. In typical strong shocks, it is given by $\gamma_c = \frac{9 m_e (1+z)}{128 m_p \sigma_T \epsilon_B n_0 c \Gamma^3 t}$ \cite{dermer}. Here $\epsilon_B$  is a fixed fraction of the magnetic energy density and the downstream energy density of the fluid. Since the 
magnetic energy density is always smaller or equal to the downstream energy density of the shocked fluid, it will be less than or equal to one. Similarly $n_0$ is the fractional number density of the electrons in the post shock region. 
Substituting the values of $\sigma_T$ (Thompson coefficient) and the redshift values that we typically have as well as the other parameters, we find that the 
$\gamma_c << 1$. This means that dominant mechanism for the energy loss of the 
electrons in the shocked fluid will be due to the synchrotron radiation.

Based on this we can now find the maximum Lorentz factor for the accelerated electron distribution. The shortest acceleration time scale expected in Fermi processes is given by the Larmor radius $r_L/c$. The Larmor radius is given by 
$r_L = \frac{\gamma m_e c^2}{e B}$. For the critical value of the magnetic field, the Larmor radius is smaller than the width of the shock generated behind the cosmic string. 
For magnetized shocks the value of $\gamma_{max}$ is given by \cite{dermer},   
\begin{equation}
\gamma_{max} = \frac{1.2 \times 10^8 \tilde{\epsilon}}{\sqrt{{B}}} \label{gamma_max}
\end{equation}
Here $\tilde{\epsilon}$ is the fractional amount of energy gained by the electron 
due to the Fermi processes. This will be less than equal to one; ${B}$ is the magnetic field which is taken to be $1$ Gauss.

In the next section, we present our model for determining the synchrotron radiation from the wakes of these Abelian Higgs cosmic strings. From the details of the electrons accelerated in the shocks of the cosmic string, we obtain the spectrum of the synchrotron radiation emitted by these electrons while crossing the magnetized wake of an Abelian Higgs string.

\section{Synchrotron radiation from electrons in the wake}
The mechanism of emission of synchrotron radiation for the case of the superconducting cosmic string depends on the current 
flowing along the cosmic string. A magnetic field is generated as the superconducting cosmic string moves through the plasma. This is given by $B(r) = 2i/cr$ where $r$ is the radial distance from the core of the string. In ref \cite{rogozin}, it was assumed
that the electrons develop a power law energy distribution in the post-shock region in the wake of these moving strings. The 
synchrotron radiation was then calculated based on the tension of the superconducting string which depended on the current flowing through the string. However, in the case of the Abelian Higgs cosmic string the magnetic field is generated in a completely different way. It is generated  due to the misalignment of density inhomogeneities and the 
temperature gradient of the shock wave by the Biermann mechanism. There are therefore two magnetic fields generated at the two sides of the shock, each directed opposite to the other. The magnetic field lines are therefore parallel to the direction of motion of the cosmic string. So the synchrotron radiation is generated by the electrons in the overdensity which move across the wake like structure.

As described in the previous section, the electron distribution in the cosmic string wake is inhomogeneous. There are regions which have a higher density of electrons and regions which have a lower density of electrons. Since these electrons are accelerated by the motion of the shock generated in the string's wake, they will lose energy as synchrotron radiation. For electrons, the peak frequency at which this loss happens is given by, 
\begin{equation}
\nu_{c} \sim \frac{eB \gamma^2}{2 \pi m c } \label{nu_c}
\end{equation}
Here the magnetic field $B$ is considered to be homogeneous over the mean free path of the electrons. 
The average power that is radiated by the electrons is given by, 
\begin{equation}
 \langle P^{syn}(\nu) \rangle  = \frac{\sqrt{3}e^3 B}{m_e c^2} \int_1^{\infty} d \gamma N_e(\gamma) R(\alpha) \label{average power}
\end{equation}
Here, $\gamma$ is the relativistic Lorentz factor of the electrons in the shock and $\alpha = \frac{\nu}{\nu_c}$. $N_e(\gamma)$ is the distribution of electrons which are moving across the cosmic string wake, $R(\alpha)$ is a function of a combination of Bessel functions.

One of the simplest models used for obtaining the spectrum is the standard one zone model. Generally, it is used for a spherical geometry but this model has also been adapted to model jets from different sources. It is also referred to in the literature as the Blob model. We will however use the geometry that is used for streaming jets. The narrow wakes of cosmic strings have strong shocks, as described in the previous section, the accelerated electrons in these shocks will resemble jets  The electrons in the jet have a Lorentz factor distribution given by $N_e (\gamma')$, where $\gamma'$ is the comoving Lorentz factor. For this model, the synchrotron radiation is given by, 
\begin{equation}
\nu F_{\nu}^{syn} = \frac{\delta_D^4 \nu'\langle P^{syn}(\nu') \rangle  }{4 \pi d^2_L} \label{flux}
\end{equation}
The Lorentz factor determines the Doppler factor for the relativistic outflow given by $\delta_D$ defined by, 
\begin{equation}
\delta_D = \frac{1}{\gamma(1- \beta cos \theta)} \label{dopple factor}
\end{equation}
Here $\beta = \frac{v_s}{c}$ and $\theta$ is the viewing angle which is very small. The other input, that is important for an observable is the luminosity distance. This distance will depend on the redshift of the object. The luminosity distance $d_L$ is given by, 
\begin{equation}
d_L = d_A (1+z)^2 \label{luminosity distance}
\end{equation}
where $z$ is the redshift and $d_A$ is the angular size distance. It is therefore the ratio of the transverse extent of the object and the angle it subtends in the sky.
The python module AGNPY \cite{agnpy} has been developed based on the one zone model and has been tested for different cases. As is seen the primary inputs come from the particular source being studied and the final output spectrum would depend upon the electron distribution, the Lorentz $\gamma$ factor of the electron distribution, the Doppler factor etc.

As mentioned before, when studying synchrotron radiation from blackholes 
and other objects, jets of plasma get ejected. These are 
 modelled using the one zone model. For the jets, the radius is no longer a constant value as they are 
scaled by the redshift parameter. The region resembles a segment of a circle with a radius that is larger than the arc that is being swept out. Having an idea of the mean magnetic field and the energy density of the electrons gives us the power emitted due to the 
radiation. The jets from the plasma outflow from the cosmic string wake can be modeled by the one zone model considering that the magnetic energy density in the plasma is of the order of $1$ Gauss \cite{dermer}. Due to this constraint, we have taken the magnetic field value to be a constant.    
   
We have modified the geometry in the AGNPY module to include the scaling due to the redshift parameter to obtain the synchrotron radiation in our case. This geometry is known as the one zone leptonic model and has been used for jets from blazers \cite{tiffany} previously.  
The first contribution of the cosmic string will come from the calculation of $d_L$. We have taken the value of $d_A$ from equation 12 to be the thickness of the wake in the comoving frame \cite{dacunha1}. Therefore, 
\begin{equation}
d_A = \frac{24 \pi}{5}(G \mu) v_s \gamma t_0 \frac{\sqrt{(1+z_i)}}{(1+z)}
\end{equation} 
Here $G \mu = 10^{-7}$,
$z_i$ is the redshift at which the wakes are formed, we consider this to be at matter-radiation equality ( $t_{eq}$). So, $z_i=z_{eq} = 3400$. The synchrotron observations are made at different $z$ which are $z = 30$, $z = 1$, $z = 0.069$. These are plotted in the final graph (figure 1).

We have considered the electron distribution in the cosmic string wake as a power law distribution. Shocks are generated in the wake of the cosmic strings \cite{stebbins,nayak} and electron distribution in shock accelerated plasma  have a power law distribution \cite{kirk}. The electron distribution is therefore  given by,    
\begin{equation}
N_e(\gamma) = K_e \frac{1}{4 \pi} \gamma^{-p} H(\gamma; \gamma_1, \gamma_2) 
\label{electron distribution}
\end{equation} 
Here, $K_e$ is the normalization constant and $p$ is the power law exponent which is close to $2$ but always slightly greater than $2$. We have studied both $p = 2.1$ and $p=2.2$. The overall results are similar and hence we present the results for $p = 2.1$. The normalization constant is calculated from the total energy of the system due to the shock acceleration. It is given by, 
\begin{equation}
K_e = \frac{(p-2)u_e}{m_e \left(\gamma_1^{(2-p)}-\gamma_2^{(2-p)}\right)} 
\label{normalization constant K_e}
\end{equation}
The function $H(\gamma; \gamma_1, \gamma_2)$ is the Heaviside function. 
The quantity $\gamma_1$ is the minimum Lorentz factor of the electrons and $\gamma_2$ is the maximum possible Lorentz factor that the electrons can achieve, $u_e$ is the total energy density of the electrons. It is given by, 
\begin{equation}
u_e = m_e c^2 \int_{\gamma_1}^{\gamma_2} \gamma d\gamma n_e(\gamma) 
\end{equation} 

In the AGNPY module, $K_e$ is calculated from the total energy of the electrons $W_e$. To calculate $W_e$, we obtain the number density of the electrons in the wake region and multiply it with the volume of the wake region.   
In the AGNPY, the volume is taken to be that of a sphere of uniform radius. We have changed the volume to that of an approximate triangular prism shape taking into account the three dimensions of the cosmic string wake. So in comoving coordinates, the volume is given by, 
\begin{equation}
V_b = c_1 \frac{t_0}{\sqrt{(1+z_i)}} \times v_s \gamma_s \frac{t_0}{\sqrt{(1+z_i)}} \times \frac{24 \pi}{5}(G \mu) v_s \gamma t_0 \frac{\sqrt{(1+z_i)}}{(1+z)}
\end{equation} 
If the cosmic string wake is generated at $t = t_i$ and the synchrotron radiation is being emitted at time $t$, then the number density of the electrons in the wake region is given by, \cite{branden}
\begin{equation}
n_e (t,t_i) =  f \rho_B(t_i) m_p^{-1} \left(\frac{(z(t)+1)}{(z(t_i) +1)} \right)^{3}
\end{equation} 
Here $f$ is the ionization fraction and $\rho_B$ is the energy density in the baryons. $\rho_B$ can be calculated from the critical density of baryons. We consider the wake to have been formed at $z = 3400$. The $N_e$ depends on the redshift parameter at which observations are being made and the total energy of the electrons is  obtained by multiplying the total number of electrons with the energy of a single electron. So for different values of the redshift, this total energy is different. For $z = 30$ , $W_e = 3.5 \times 10^{53}erg$, for $z = 1$ , $W_e = 2 \times 10^{52} erg$, and for $z = 0.069$ , $W_e = 10^{52} erg$.  $W_e$ is the input for the AGNPY module.   As the $W_e$ is different for different redshifts $z$, hence the spectrum is different for different values of $z$. We have therefore plotted the synchrotron spectrum for different values of $z$ in figure 1.   
  
The synchrotron radiation emitted by the electrons for an isotropic pitch angle distribution is given by, 
\begin{equation}
\nu F^{syn}_{\nu} (p) = \frac{3^{(p+2)/2}}{2^{(p+1)/2}} a(p) \frac{4}{3} c \sigma_T U_B K_{e} \left(\frac{\nu}{\nu_B} \right)^{(3-p)/2} \label{flux(z)}
\end{equation}
Here $\sigma_T$ is the Thomson cross section, $U_B = \frac{B^2}{8 \pi}$,  $\nu_B = \frac{e B}{2 \pi m_e c}$. As mentioned before, for relativistic shocks $p \approx 2 $ and $a(p) = 0.1032 $. The $U_B$ considered in our final calculations corresponds to the critical magnetic energy.    
Details of these calculations are available in ref.\cite{dermer}.

In the previous section, we have discussed that the cosmic string moves with velocities close to that of light, from reference \cite{sornborger2}, we see that the velocity of the string is taken to be approximately $0.5$ times the speed of light. In the previous section we have found that the minimum Lorentz factor $\gamma_{min}$ is one. Again as discussed in the previous section, the electrons are also accelerated by reflection of ions from the shock front. Substituting the 
values of magnetic field and $\tilde{\epsilon} \le 1$, 
$\gamma_{max} \le 10^8$.  So we have taken the maximum value of the Lorentz factor to be $\gamma_{max} = 10^{7}$. We mention here that the Lorentz factor corresponding to the cosmic string $\Gamma$ is considered to be one.  The value of $\delta_D$ depends on the Lorentz factor. $\delta_D$ is an important parameter in the emission spectrum of the synchrotron radiation. It also depends on the redshift $z$. We have used this as a parameter and used different values of redshift to obtain the final spectrum. Since radiation has been observed from very low redshift objects only, we keep the maximum of $z$ as $30$.

\section{Results}

\begin{figure}
	\includegraphics[width = 0.7\linewidth]{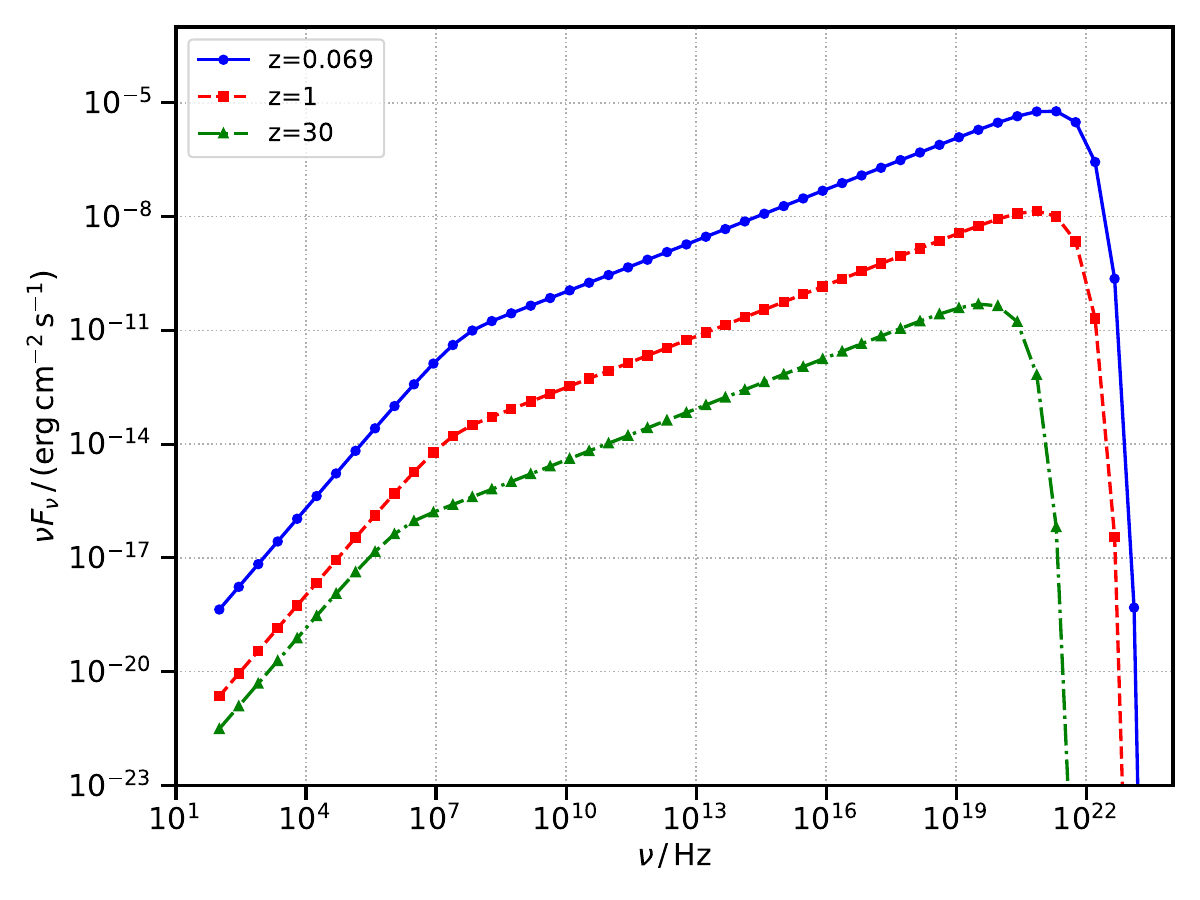}
	\caption{The synchrotron spectrum for different values of the redshift parameter at $\Gamma = 1$ and $p = 2.1$.}
	\label{fig:redshift}
\end{figure}

We now present the synchrotron spectrum for the different parameters discussed in the previous section. We plot the synchrotron spectrum for the relativistic string ($\Gamma \approx 1 $) for various redshift values. The redshift values are limited by the observational component and so as mentioned before, the largest redshift we have considered is $z=30$ while the smallest is $z = 0.069$. The value of $p = 2.1$. The plots are shown in fig \ref{fig:redshift}. 
We see that the turnover frequency is about $10^{7}$ GHz depending on the redshift value. The spectral break comes after $10^{21} -10^{22} $ GHz. As expected we find that, the spectral break shifts to low frequencies as we move back in time. The synchrotron spectrum obtained can be fit with three distinct power laws. The initial low frequency self absorption part has a power law exponent of $\sim 1.28$, this is followed by the optically thin emission region which has a very low spectral index ($\sim 0.42$). This is the region of the spectrum where the relativistic electrons are scattered multiple times. Finally it dips very sharply in the high frequency range. This is different from the usual spectrum obtained from the Gamma Ray Bursts where there is a more bell shaped curve.

Galactic radiation has been observed at such frequencies in several cases. Due to the high sensitivity of the all sky surveys there are many sources of radiation which have been detected. Not all these sources can be associated definitely and unambiguously with known sources such as an Active Galactic Nuclei (AGN). There are some unidentified sources which are currently being studied \cite{hannes}.  We proceeded to search the catalogues to determine if there are any unidentified sources in the frequency range of the synchrotron spectrum that we have obtained from cosmic string wakes. 
In the infra red (IR) region, the Wide-field Infrared Survey Explorer (WISE) data has several point sources attributed to Blazers which are also in the range of the spectrum that we have obtained. The similarity to blazars comes from the fact that the the gamma ray emission in blazars has a very similar geometry to the expected gamma ray emission from cosmic string wakes. Also the emission in blazers may be due to shock-shock collision, magnetic reconnection or turbulence, all three of which can occur in the cosmic string wakes too. Thus it would be difficult to distinguish between the two sources; a blazer and a cosmic string wake.
One point which distinguishes the two spectra is that the blazer usually has two broadly peaked components whereas the cosmic string wake will have only one broad peak. 
     
\begin{figure}
	\includegraphics[width = 0.7\linewidth]{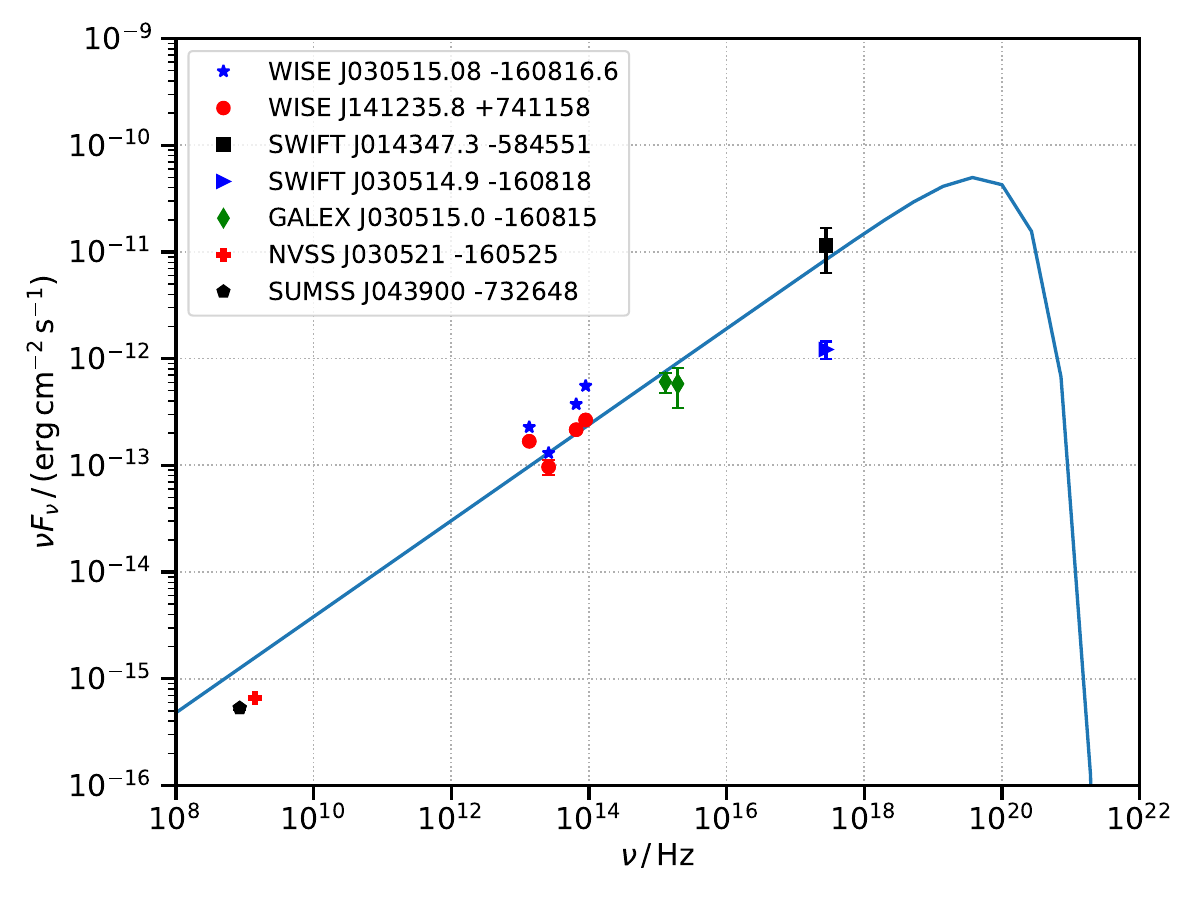}
	\caption{The synchrotron spectrum with data points from the WISE, SWIFT, GALEX NVSS and SUMSS catalogues. The spectrum is obtained for $z = 30$ with other parameters remaining the same.}
	\label{fig:wise}
\end{figure}

We now proceed to plot a few sources from the various observations in fig. 2 to show that the current spectra that we have obtained is within the scope of current observations. We ran single object searches in different catalogs (WISE, SWIFT, GALEX, NVSS and SUMSS)  and found that  there are some unidentified sources in the frequency range of the synchrotron spectrum we have obtained. The frequency range is from $10^{8} - 10^{22} Hz$ and the flux ranges from $10^{-15} - 10^{-11} erg cm^{-2} sec^{-1}$.  We used the WISE catalogue to search for sources of infrared radiation in this flux scale. In fig \ref{fig:wise}, we have plotted a few points from the WISE catalogue \cite{WISE} on the obtained spectrum for $\Gamma = 1$. Though,
not a great fit, some of the data points are close to the obtained spectrum. 
Similarly, we have included observed fluxes from the SWIFT \cite{SWIFT},  the GALEX  \cite{GALEX}, the NVSS \cite{NVSS} and the SUMSS \cite{sumss} catalogues.  We have used the data that is available online and find that different surveys in different frequency bands have observed fluxes in this range. This seems to indicate that a more detailed study of synchrotron radiation in cosmic string wakes may lead to results matching with experimental data. We have not included the position of the sources as that does not contribute anything more to our analysis. However it is possible to do a more detailed analysis by including information about the position of the sources and other observational features in the analysis. The reason we are interested in studying the data from various unknown sources is because it seems that the best way to identify these string wakes is to coordinate mulitple signals from density inhomogeneities and electromagnetic radiation to improve the chances of identifing these objects. However, currently we do not have access to the large scale network models which can lead to such coordination. We would like to improve the predictions from synchrotron radiation in the future and plan to look at it in more detail.   

\section{Comparison with the synchrotron spectrum from superconducting strings}

The synchrotron spectrum of electrons moving close to a superconducting string has been studied in detail previously \cite{rogozin}. There are however quite a lot of differences in the calculation of the spectrum between the case of the superconducting strings and loops and the case of the Abelian Higgs strings that we have presented. It is the current in the superconducting string that generates the magnetic field through which the electrons move in the superconducting string case hence the current $i$ is an important parameter in the final spectrum for the case of the superconducting string. In the case that has been discussed here, there is no string current involved in the generation of the magnetic field. The only cosmic string parameter that is relevant here is the deficit angle in the space time of the cosmic string. The cosmic string wake has a narrow sheet like geometry. Strong shocks are generated in the cosmic string wakes. Electrons moving across these wakes are scattered by ions reflected by the shock. This 
accelerates the electrons to very high energies. This energy is then radiated as 
synchrotron radiation.  

The obtained spectrum for both the superconducting string and the Abelian Higgs string is over a large range of frequencies starting from $10^2$ Hz and going all the way upto $10^{23}$ Hz. We now make a brief comparison of the results obtained from superconducting strings as in ref. \cite{rogozin} and the results we have obtained for the Abelian Higgs strings. In ref. \cite{rogozin} we see that the synchrotron radiation has a range starting from $10^6$ Hz and going all the way upto $10^{23}$ Hz. The flux has a range of $10^{-24}$ to $10^{-14}$ for the superconducting string spectrum. For the same range of frequencies, we get the flux between $10^{-20}$ to $10^{-5}$ for the redshift parameter $z = 0.069$. The peak in the energy flux is higher in the current case. Even for the other values of $z$, we have studied, the peak flux is higher. The spectrum where different points have been mapped on the spectrum is for $z = 30$ and there we see that the peak is around $10^{-10}$, so there is a greater possibility of detecting synchrotron radiation from electrons moving in the wake of a cosmic string where the magnetic field is formed from the Biermann mechanism.

The nature of the spectrum is also different in the two cases for the lower frequencies depending on the value of the redshift. The low frequency self absorption part is not so prominent in the case of the superconducting strings. However, it is quite significant for the case of the shock accelerated synchrotron radiation emitted from wakes where the magnetic field is generated by the  Biermann mechanism. Thus there are several features which distinguish the spectrum of the superconducting string from the spectrum that we have obtained.

\section{Summary and Conclusions}
In this work we have studied the synchrotron radiation emitted by relativistic electrons moving in a cosmic string wake. The narrow width of the wake gives it a thin sheet like structure. The magnetic field is generated by the Biermann mechanism. The string is not a superconducting cosmic string and hence the magnetic field calculation in the wake region is also different. In the case of the superconducting cosmic string, the synchrotron spectrum depends on the current $i$ flowing along the string. Here the shocks generated in the string wakes are the
primary factors determining the synchrotron radiation. The spectrum of
radiation obtained is also different from the case of the superconducting string. The flux of radiation obtained is higher in this case and hence there is greater chance of detection of this radiation.

We obtain the spectrum for different values of the redshift $z$ using a one zone leptonic model with a modified geometry. Unlike the Gamma Ray Bursts studied previously the $\gamma$ factor is limited to lower values in this case. Electrons moving in a cosmic string wake can generate $\gamma$ factors ranging from $1 - 10^{7}$. The lower limit comes from the particles moving into the wake due to the conical space time of the moving string. For the higher limits, the electrons are the shock accelerated electrons. 
We have assumed that the overall magnetic field is homogeneous over the width of the wake. Though it is possible that the magnetic field will have some small scale fluctuations due to turbulence in the plasma, for the current work we have not looked at such details.

We have found that the frequency spectrum obtained from the relativistic  electrons moving in the wake of a cosmic string will be over a wide range of frequencies. There is a broad region in the spectrum due to the multiple scattering of the electrons. The spectrum can be fitted with three separate power law exponents. There is only one broad peak in the spectrum. Thus the nature of the spectrum differs from the spectrum usually obtained from blazers or AGN's. Since there are still quite a few unidentified sources of radiations in the catalogues for the all sky surveys, we have attempted to find some of the data for the point sources in the given frequency range. We do find that several of the surveys (WISE, SWIFT, GALEX, NVSS and SUMSS) have unknown sources with similar flux in the range of frequencies that are covered by the cosmic string wake. 

There are some limitations to this study. We have studied the synchrotron radiation emitted by electrons in a single long cosmic string wake. Unfortunately, the way cosmic strings evolve this may not be the scenario. Cosmic strings come close together and can have overlapping wake structure, also the time at which the strings are formed are different and we have only considered strings formed at a particular time. We plan to extend this study further to include more details about the cosmic string wake in the modeling of the synchrotron radiation.
 
Currently we have used a rather simple one zone leptonic model which is generally used for jets and simplifies the plasma in the wake to a homogeneous plasma. We have also used a homogeneous magnetic field in our calculation which may not be the case for the magnetic field in the cosmic string wake. It is quite possible that after generation  of the seed field in the wake of the cosmic string, the evolution of the field by turbulence leads to a randomized field over short lengthscales. This would change the final spectrum of the synchrotron radiation. We hope to look at all these details in a future work so that we have a more distinct picture of the synchrotron radiation from electrons moving across the cosmic string wakes.

\begin{center}
 Acknowledgments
\end{center}  

This research was carried out on the computational facility  set up from funds given by the DST- SERB Power Grant no. SPG/2021/002228 of the Government of India. S.N acknowledges financial support from CSIR fellowship No. 09/414(2001)/2019-EMR-I  given by the Human Resource Development, Government of India. D.K acknowledges financial support from DST- SERB Power Grant no. SPG/2021/002228 of the Government of India. The authors would like to thank Abhisek Saha for help with the numerical codes and Sayantan Bhattacharya and Susmita Barman for suggestions and discussions regarding the experimental results. We also thank the anonymous referees for their comments which has helped to improve this work considerably.

\end{document}